\documentclass[12pt,preprint]{aastex}

\slugcomment{ApJ, in press}

\shorttitle{Early-type galaxies in the Coma cluster}
\shortauthors{Poggianti et al.}

\begin{document}

\title{Ages of S0 and elliptical galaxies in the Coma cluster
\footnote{Based on observations made with the William Herschel Telescope 
operated on the island of La Palma by the Isaac Newton Group in the 
Spanish Observatorio del Roque de los Muchachos of the Instituto de 
Astrofisica de Canarias.}}
\author{Bianca M.\ Poggianti,$^{\!}$\altaffilmark{2}
Terry J.\ Bridges,$^{\!}$\altaffilmark{3}
Dave Carter,$^{\!}$\altaffilmark{4}
Bahram Mobasher,$^{\!}$\altaffilmark{5}
M. Doi,$^{\!}$\altaffilmark{6} 
M. Iye,$^{\!}$\altaffilmark{7} 
N. Kashikawa,$^{\!}$\altaffilmark{7} 
Y. Komiyama,$^{\!}$\altaffilmark{8}  
S. Okamura,$^{\!}$\altaffilmark{9,11} 
M. Sekiguchi,$^{\!}$\altaffilmark{10}
K. Shimasaku,$^{\!}$\altaffilmark{9}  
M. Yagi,$^{\!}$\altaffilmark{7}
N. Yasuda$^{\!}$\altaffilmark{7}
}
\smallskip

\affil{\scriptsize 2) Osservatorio Astronomico di Padova, vicolo dell'Osservatorio 5, 35122 Padova, Italy, poggianti@pd.astro.it}
\affil{\scriptsize 3) Anglo-Australian Observatory, PO Box 296, Epping, NSW 2121, Australia}
\affil{\scriptsize 4) Liverpool John Moores University, Astrophysics Research Institute, Twelve Quays House, Egerton Wharf, Birkenhead, Wirral, CH41 1LD, UK}
\affil{\scriptsize 5) Space Telescope Science Institute, 3700 San Martin Drive, Baltimore MD 21218, USA \\
Affiliated with the Space Sciences Department of the European Space Agency}
\affil{\scriptsize 6) Institute of Astronomy, School of Science, University of Tokyo, Mitaka, 181-0015, Japan}
\affil{\scriptsize 7) National Astronomical Observatory, Mitaka, Tokyo, 181-8588 Japan}
\affil{\scriptsize 8) Subaru Telescope, 650 North Aohoku Place, Hilo, HI 96720, USA}
\affil{\scriptsize 9) Department of Astronomy, University of Tokyo,
Bunkyo-ku, Tokyo 113-0033, Japan}
\affil{\scriptsize 10) Institute for Cosmic Ray Research, University of Tokyo,
Kashiwa, Chiba 277-8582, Japan}
\affil{\scriptsize 11) Research Center for the Early Universe, School of Science, University of Tokyo, Tokyo 113-0033, Japan}

\begin{abstract}
The ages of stellar populations in 52 elliptical and S0 galaxies in
the Coma cluster are investigated using a new spectroscopic sample.
More than 40\% of the S0s are found to have experienced
recent star formation in their central regions during the last $\sim
5$ Gyrs, while such activity is absent in the ellipticals.  
Galaxies in this sample have absolute magnitudes in the range $-20.5 < M_B <
-17.5$, and the
fraction of S0 galaxies with recent star formation is higher at
fainter luminosities.  The observed luminosity range of S0 galaxies
with signs of recent star formation activity is consistent with them
being the descendants of typical star-forming spirals at intermediate
redshift whose star formation has been halted as a consequence of the
dense environment.
\end{abstract}


\section{Introduction}
The formation mode and epoch of spheroids are among the most
important unsolved issues in observational cosmology.  These questions
involve two possibly distinct aspects: when did the stars in spheroids
form (i.e. their star formation history), and when were these stars
assembled in a galactic structure that can be classified as a
spheroid?

The bulk of stars in spheroid-dominated (early-type) galaxies are
believed to be old, at least in clusters
\citep{bow92,ell97,kel97,kod98,sta98,kel01}, but the frequency and
amount of later star formation activity is still a matter of debate.
To discriminate among different formation scenarios, it is crucial to
determine whether field spheroidal galaxies had more extended star
formation histories than their cluster counterparts (for opposite
views see \citet{ber98} and \citet{men99,men01}, also compare Renzini
1999 vs. Gonz\'ales 1993 and Trager et al. 2000).

In most studies, early-type galaxies (ellipticals and S0s) are treated
as one class of objects, due to the general similarity in their colors
and their ``vicinity'' along the Hubble sequence.  This is in spite of
the fact that also remarkably different formation processes have been
proposed for ellipticals and lenticulars.  In the case of ellipticals,
the most cited alternatives are formation at high redshift through
dissipationless collapse (the so-called monolithic scenario,
e.g. Larson 1974), or a merger of disk systems (in the framework of
hierarchical galaxy formation models, e.g.  Baugh et al. 1996,
Kauffmann 1996).  Concerning the formation of S0 galaxies, it is a
longstanding issue whether they are ``primordial'' objects (i.e. they
formed as S0s), or whether they were converted from spirals which lost
their gas supply either by galaxy-galaxy collisions, ram-pressure
stripping or gas evaporation in the hot intracluster medium, galactic
winds or loss of their gas-rich envelopes \citep{spi51, gun72, fab76,
cow77, bur79, dre80a, lar80}.

A spiral origin for S0 galaxies seemed to be favored by the discovery
of large numbers of blue (presumably spiral) galaxies in clusters at
$z>0.2$ \citep{but78,but84}, coupled with the abundance of S0s and
ellipticals in clusters in the local universe \citep{hub31,oem74}.  A
renewed interest in studying the stellar population ages of
ellipticals and S0 galaxies separately arose in recent years, when the
high spatial resolution imaging achieved with the {\it Hubble Space
Telescope} (HST) uncovered a strong {\it morphological} evolution
taking place in rich clusters during the last few Gyrs. There is an
overabundance of spirals in cluster cores at z=0.5 and the S0 galaxies
are proportionally (a factor 2 to 3) {\it less} abundant than in
nearby clusters, while the fraction of ellipticals at z=0.5 is the
same or larger than at z=0 \citep{dre97}.  The progression of this
morphological evolution between z=0.5 and now has been recently
clarified by \citet{fas00}, who explored for the first time the
intermediate-redshift range z=0.1-0.25.  These results strongly
suggest that a large number of cluster spirals observed in distant
clusters have evolved into the S0s that dominate rich clusters today
(see also Kodama and Smail 2001).

Observationally, the existence of systematic differences between the
ages of stellar populations in ellipticals and lenticulars is still
controversial.  Neither photometry nor spectroscopy of early-type
galaxies in high-redshift clusters has been able to reveal any
statistically significant difference \citep{ell97,jon00}, and the same
is true for other works on lower redshift clusters
\citep{jor97,jor99,zie01,tra01}.

However, other studies have found evidence for young and intermediate
age stellar populations in a significant fraction of lenticulars
\citep{cal83,gre89,bot90,fis96}.  Recently, two studies have
uncovered a difference in the recent star formation histories of
cluster S0 and elliptical galaxies.  A high-quality line strength
analysis has shown that most of the (faint) lenticular galaxies in the
Fornax cluster have lower luminosity-weighted ages than the (brighter)
ellipticals \citep{kun98,kun00}.  In Abell 2218 at z=0.17,
high-precision optical and near-IR photometry has found that
ellipticals at all magnitudes and luminous S0s are old and coeval,
while the faintest S0s appear to have younger luminosity-weighted ages
\citep{sma01}. 

Both of these studies find that the differences between Es and
S0s mostly stand out at faint magnitudes. The luminosity range
explored by the different studies could be the key to understanding
the reason for the apparently contrasting results found so far
\citep{sma01}, but the relative role played by mass and morphology (S0
vs E) in driving the evolutionary history of cluster early-type
galaxies is still unclear \citep{kun98,zie01}.

The central issue is whether recent star formation in a significant
fraction of the lenticulars is a widespread phenomenon in clusters.
Additionally, it would be important to conclusively understand if and,
eventually, why this effect is evident only in a certain luminosity
range.

To address these questions we present here a study of the
spectroscopically-determined ages and metallicities of ellipticals and
lenticulars in what is considered the local prototype of rich
clusters, the Coma cluster.  Our sample comprises 52 galaxies 
(19 ellipticals and 33 S0s) covering
a broad range in luminosity ($M_B \sim -20.5$ to $M_B \sim
-17.5$).  We will show that the populations of S0s and Es differ for
the presence/lack of recent star formation activity in a significant
fraction of these galaxies. We also discuss the luminosity dependence
of these findings and its possible origin.

\section{Observations and analysis}

This work is based on the spectroscopic survey of galaxies in the Coma
cluster presented in \citet[Paper II]{mob01}.  Multi-fiber spectra
with a resolution 6--9 \AA $\,$ and a signal-to-noise
$\sim 15-19$ were obtained for a random subset of
galaxies from the sample of \citet{col96}\footnote{Complemented by an
unpublished list of redshifts kindly provided by M.~Colless.}  in two
areas of $\sim 1\times 1.5$ Mpc towards the center and the South-West
region of the cluster. A full description of the sample selection,
the observations and data reduction can be found in Paper II, where it
is shown that this is essentially a magnitude limited sample with no
significant bias.

In this paper we present the stellar population properties of
the elliptical (E) and S0 galaxies.  Morphological classifications
from \citet{dre80b} are available for 77 galaxies in the Mobasher et
al. sample, of which 19 are Es and 33 are S0s.  All except two of
these galaxies are in the magnitude range $13< R < 16$ ($-20.5 < M_B <
-17.5$).  Intermediate types such as E/S0 and S0/E (2 galaxies) and
SB0s (3 galaxies) have been excluded from the present analysis.
Dressler classified as S0s those galaxies with a clearly recognizable
non-spheroidal (disk or lens) component. When viewed face-on, they
display an intensity discontinuity between the bulge and the disk.
Galaxies with a smooth radial profile, with no intensity
discontinuities, were classified as ellipticals. Disk galaxies with a
clear spiral or outer ring pattern were classified as spirals.

Line indices of the Lick/IDS system and emission line equivalent
widths were measured from the spectra and compared with
spectrophotometric models to derive luminosity weighted ages and
metallicities as described in \citet[Paper III]{pog01}.  The line
indices used in this paper, as well as those of the whole
spectroscopic sample, will be published in a later paper of the
series. Great care was taken to ensure an accurate calibration onto
the Lick/IDS line index system, correcting for the spectral
resolution, the galaxy velocity dispersion and any residual offset.
The comparison with standard Lick stars yields an uncertainty of $\pm
0.1$ \AA $\,$ in our calibration of the $\rm H\beta$ index (see Paper
III). As we will show below, this is significantly smaller than the
random errors on our $\rm H\beta$ measurements, hence the latter
dominate the age uncertainty.  Emission-line spectra were excluded
from the analysis of the Lick indices, and no emission correction was
applied to the index measurements.

To derive the stellar population properties, 
we have used the Padova version of Worthey's
models (see Paper III), which includes an accurate
treatment of the horizontal branch of low metallicity stars. The main
differences between the Padova and the standard Worthey models in the
$\rm H\beta$-$\rm Mg_2$ diagram that is employed below can be
summarized as follows: a) the standard version does not cover the
range [Fe/H]$< -0.225$ for ages $< 8$ Gyr, and b) the $\rm H\beta$
strengths at ages 5-12 Gyr are generally lower in the Padova version,
translating into age differences of typically 2 Gyr. At ages less than
4 Gyr, the difference is $\sim 0.5$ Gyr. We note that,
while \it absolute ages \rm are highly uncertain in this type of
models, \it relative ages \rm are much less affected by the model
uncertainties. The current sample mostly consists of
intermediate-luminosity galaxies, which do not display a high [Mg/Fe]
ratio as more luminous ellipticals do \citep[see also Paper
III]{wor92,tra00,kun01}, therefore no attempt has been made to vary
the abundance ratios of the models.

We point out that: a) the spectra refer to the central 2.7 arcsec (1.3
kpc) of each galaxy, hence recent star formation episodes in the outer
regions would not be detected; b) the quantities derived from the
spectra are \sl luminosity weighted \rm ages and metallicities. Other
quantities of interest -- such as for example the mass fraction
involved in the latest star formation episode -- remain unknown; c)
absolute ages/metallicities are much more uncertain than relative
ages/metallicities, due to the uncertainties intrinsic to the models.
We assume a distance modulus to Coma of 35.16 ($<v>=7000 \, \rm km \,
s^{-1}$, $H_0=65 \, \rm km \, s^{-1} \, Mpc^{-1}$).

\section{Results}

The main result is presented in Fig.~1, showing the $\rm H\beta$
versus $\rm Mg_2$ diagram of ellipticals (empty symbols) and S0
galaxies (filled symbols). The spectrum of one S0 galaxy in our sample
has very strong emission lines indicative of a current starburst and
has been excluded from this and the following plots.  

There is a striking difference between the populations of ellipticals
and S0s.  All but one of the ellipticals are consistent within the
errors with ages $>9$ Gyr. In contrast, 13 out of 32 S0 galaxies have
luminosity weighted ages smaller than 5 Gyr.\footnote{We note that
only one of these -- the galaxy with the luminosity-weighted age $\sim
1.5$ Gyr at $\rm H\beta = 3.15$ -- is classified as a ``k+a'' galaxy
(=''E+A'') according to the MORPHs classification scheme by
\citet{dre99}. Another 3 S0s with luminosity-weighted ages less than 5
Gyrs were classified as k/k+a's (marginal k+a cases).  There is an
excellent agreement between the MORPHs classification and the results
based on the Lick indices for the strongest Balmer line cases.  As
expected, the MORPHs classification is able to identify those galaxies
with luminosity-weighted ages $< 2$ Gyr regardless of metallicity but
-- due to the line strength thresholds adopted -- it is insensitive to
older episodes of star formation which, instead, can be still detected
in the Lick system.}  Hence, more than 40\% of the S0s appear to have
experienced star formation at a recent epoch, while such widespread
recent activity is absent in the elliptical population.  These results
are corroborated by the analysis of the $\rm H\gamma^F$ -- $<$Fe$>$
diagram that provides a similar but independent method of deriving
ages and metallicities.\footnote{During the course of this work,
another spectroscopic survey of Coma galaxies has been presented
\citep{cas01}. These authors investigate the star formation history of these
galaxies using different methods, comparing with the spectral energy
distributions of synthetic models.  While a direct comparison of the
``ages'' derived would be meaningless -- due to the different meaning
of the word ``age'' between our and their analysis -- it is
interesting to note that in the great majority of cases the two
studies agree in identifying galaxies with/without recent star formation:
among the 21 galaxies in common (10 ellipticals and 11 S0s), both
studies assign ``old'' ages (lack of recent star formation) to 15
galaxies, identify 1 emission-line galaxy in common and detect recent
star formation ($\sim 3$ Gyr or less) in 3 galaxies with
absorption-line spectra. There is only 1 discrepant case of a galaxy
in which we detect a recent activity not reported in their work, and
another case where the results agree within the errors.}

Inspection of Fig.~1 shows that the metallicity range covered by this
sample is quite wide, typically from [Fe/H]=-0.7 to 0.4,
i.e. abundances from 0.2 to 2.5 times solar. A metallicity spread is
expected, given the broad magnitude range explored and the existence
of a correlation between mean metallicity and galaxy luminosity.
Ellipticals and S0s follow a broadly similar metallicity-luminosity
relation, as shown in the top panel of Fig.~2. However, the relation
for the ellipticals ($Z= 1.850 -0.137 \times R$, $\sigma=0.180$) is
slightly flatter and less scattered than the relation for the S0s ($Z=
5.721 -0.400 \times R$, $\sigma=0.284$). This could explain the finding
of \citet{fis96} that S0s might have a steeper $\rm Mg_2$-velocity
dispersion relationship than Es.

The bottom panel of Fig.~2 presents the luminosity-weighted ages as a
function of magnitude. The population of S0s with young
luminosity-weighted ages is clearly visible in the region of the plot
below the dotted line.  Only 4 out of the 13 S0 galaxies with
luminosity weighted ages $< 5$ Gyr are brighter than $R=14.5$ ($M_B
\sim -19$).  In general, the luminosity distributions of Es and S0s
are different, with the S0 distribution being more skewed towards
fainter magnitudes. In fact, the E:S0 number ratio changes from 5:6 at
$13<R<14$ to 9:12 at $14<R<15$ and 5:13 at $15<R<16$.

It is therefore important to assess how luminosity effects come into
play.  This is better exemplified in Fig.~3, where the age
distributions of Es and S0s are shown for the whole samples, and for
galaxies brighter and fainter than $R=14.5$ ($M_B \sim -19$)
separately. The total age distributions of lenticulars and ellipticals
differ significantly (with a 93.3\% probability according to a
Kolmogorov-Smirnov test). A higher proportion of
``young'' S0s in the faint subsample compared to the bright subsample
 is clearly visible in the middle panel of Fig.~3,
and the difference between the age
distributions of Es and S0s is more conspicuous at faint magnitudes.

This luminosity dependence of the evolutionary histories of S0
galaxies is naturally explained if a fraction of the S0s were
previously star-forming spiral galaxies.  We computed the luminosity
evolution of a galaxy whose star formation was truncated at some time
between 2 and 5 Gyr ago.  For galaxies with a star formation history
(prior to truncation) typical of Sa, Sb, Sc and Sd galaxies
\citep{bar97}, we find that the B-band luminosity fades by 0.5 -- 1.5
magnitudes, depending on the truncation epoch and galaxy type. In the
event of a starburst prior to the truncation, the luminosity evolution
would be greater.  Cluster spirals at intermediate redshift have
$M_V^{\star} \sim -21$ \citep{sma97} in the cosmology adopted here,
thus even in the most conservative scenario (only truncation, no
starburst, B-V=0.4)\footnote{$B-V \sim 0.4$ for a very late-type
spiral, $\sim 0.9$ for an early-type galaxy.}, the passive descendants
of spirals are expected to be typically fainter than $M_B =-20/-19$,
corresponding to $R=13.5/14.5$ in our Coma sample.

Consequently, the great majority of galaxies that evolved from the
spiral into the S0 class should not be found at the bright end of the
luminosity function. The luminous side of the S0 luminosity function
could be filled up by another population of S0s which did not evolve
from spirals (or evolved from brighter spirals at higher redshifts).  
If the morphological transformation spiral
$\Rightarrow$ S0 is accompanied by a conversion of a star-forming
galaxy into a passive one, it is not surprising that those S0s with signs
of recent star formation are preferentially lower luminosity galaxies.
This could be the reason why many previous studies -- concentrating
primarily on the most luminous galaxies -- have not found significant
differences between the evolutionary histories of S0 and elliptical
galaxies.

A question of great importance is whether it is {\it mass}, {\it morphology}, 
or {\it environment} that plays the dominant role in determining the
evolutionary history of early-type galaxies
\citep{kun98,kun00,sma01,zie01}.  The results presented here indicate
that the distinction between an S0 and an elliptical (i.e. the
presence of a disk/lens according to Dressler's classification) is
able to separate classes of galaxies with different star formation
histories. These differences are noticeable between E and S0 galaxies
{\it of similar luminosity}.  Hence, morphology appears to be the
galactic property better correlated with the presence of recent
episodes of star formation, at least in clusters like Coma. However,
we have discussed how the differences between S0s and ellipticals can
only be appreciated at $M_B > -20$, and become more and more evident
going towards fainter magnitudes.  In this respect, our findings for
the Coma cluster are similar to those by \citet{kun98} for the Fornax
cluster and by \citet{sma01} for Abell 2218. As far as galaxy
metallicity is concerned, galaxy luminosity (i.e. mass) seems to be
the primary parameter controlling it (see Paper III).

While the effects of the cluster environment may be a viable mechanism
to make spirals evolve into S0s, the existence of field lenticulars
clearly proves that this cannot be the only mechanism of S0
production. Two (or more) formation processes might be at work, of
which one could be effective only in clusters and would be responsible
for the formation of a fraction of the faint S0s.  If this is the case
and a large proportion of cluster S0s were spirals at higher redshift
as suggested by the HST studies \citep{dre97}, then the luminosity
function of S0 galaxies in distant clusters could be significantly
different from the present-day luminosity function, likely with a
lower number ratio of faint to bright S0s at high z.  For the same
reason, the {\it field} and the {\it cluster} S0 luminosity functions
at z=0 would also be expected to differ.  Studies of both the
luminosity and the age distributions of S0s in different environments
would be very valuable in placing constraints on acceptable
evolutionary scenarios.

In distant clusters, a large population of
post-starburst/post-starforming galaxies in which star formation
stopped some time during the previous 1.5 Gyr has long been known to
be present \citep[(MORPHs collaboration) and references
therein]{dre83,cou87,dre92,bar98,dre99}. It has been suggested that
these are recently infallen field galaxies whose star formation has
been suppressed by the cluster environment.  Furthermore, since most
of these post-starforming galaxies have spiral morphologies at z=0.5,
it has been proposed that at least some of these spirals are going to
evolve into S0s at a later time, and that the timescale for
morphological transformation (from spiral to S0) must be {\it longer}
than the duration of the observational signature of the cessation of
star formation (1.5 Gyr) \citep{pog99}.  These
post-starburst/post-starforming spirals would be an intermediate step
between star-forming spirals and passive S0s.  In this scenario, the
fact that in Coma we find a number of S0s with ``recent'' star
formation could appear to be at odds with the results found at z=0.5
where the post-starburst/post-starforming galaxies are mostly {\it
spirals}. This apparent inconsistency is resolved if one considers
that the Lick system employed here allows an exploration also of
luminosity-weighted ages between 2-5 Gyr, so the times elapsed since
the halting of the star formation that we can detect in this work are
much longer than the post-starburst timescales which have been
identified in distant clusters (1.5-2 Gyr at most). The S0s with
recent star formation observed here could be a later evolutionary
stage of the post-starburst/post-starforming spirals observed at
z=0.5, but not as advanced as {\it passive} S0s observed in Coma and
other clusters at z=0.

Finally, it is interesting to examine the distribution in the sky of
galaxies of different Hubble types and ages (Fig.~4).  In our sample
the phenomenon of recent star formation in some lenticulars (filled
symbols in the figure) is {\it not} preferentially observed in the
South-West region around NGC 4839\footnote{See also Caldwell and Rose
(1998) who found a similar result in their sample of low-luminosity
early-type galaxies in Coma.}. This seems to be in contrast with the
findings of \citet{cal93,cal97}, who argued for a higher incidence of
galaxies with recent star formation in the South-West area than
towards the central region, and proposed that this phenomenon could be
mostly related to the (possibly infalling) NGC 4839 group. Again, it
is possible that the current study is sensitive to a wider
luminosity-weighted age range than the studies by Caldwell and
collaborators, and therefore identifies also star formation episodes
that temporally preceded the last activity detected by these authors.
The different epochs of activity could be associated with groups of
galaxies in different regions of the cluster.

Fig.~4 also shows that S0 galaxies with luminosity-weighted ages $< 5$
Gyr have an asymmetric distribution with respect to the center of
Coma, being preferentially located in a region east/north-east of
NGC4874 (identified by the letter ``A'' in Fig.~4). Instead, older S0
galaxies and ellipticals (triangles and circles with a central dot,
respectively) are found to be spread out both north and south of
NGC4874. This perception by eye of a difference in the spatial
distribution of ``young'' and ``old'' S0s is not confirmed by a 2D
Kolmogorov-Smirnov test, which turns out to be inconclusive (the
probability is only 32.8\%).  Clearly, larger samples are needed for
reaching statistically significant conclusions.  The possibly higher
level of clumpiness of the S0s with recent star formation could be due
to the presence of substructure in Coma 
\citep[][who found an apparently bi- or trimodal 
velocity distribution for Coma S0s]{biv96,col96,zab93}, possibly being
related to the NGC4889 merger event. At least some
of the ``young'' S0s in the east/north-east region could be part of
the same group, have accreted onto the cluster approximately at the
same epoch, and consequently had their star formation halted by the
effects of the cluster on similar timescales, as suggested by the fact
they have comparable luminosity-weighted ages in Fig.~1.

\acknowledgments
We are grateful to Matthew Colless, Ian Smail and Andrea Biviano
for their comments and useful discussions regarding this work.

\clearpage

\begin{figure}
\plotone{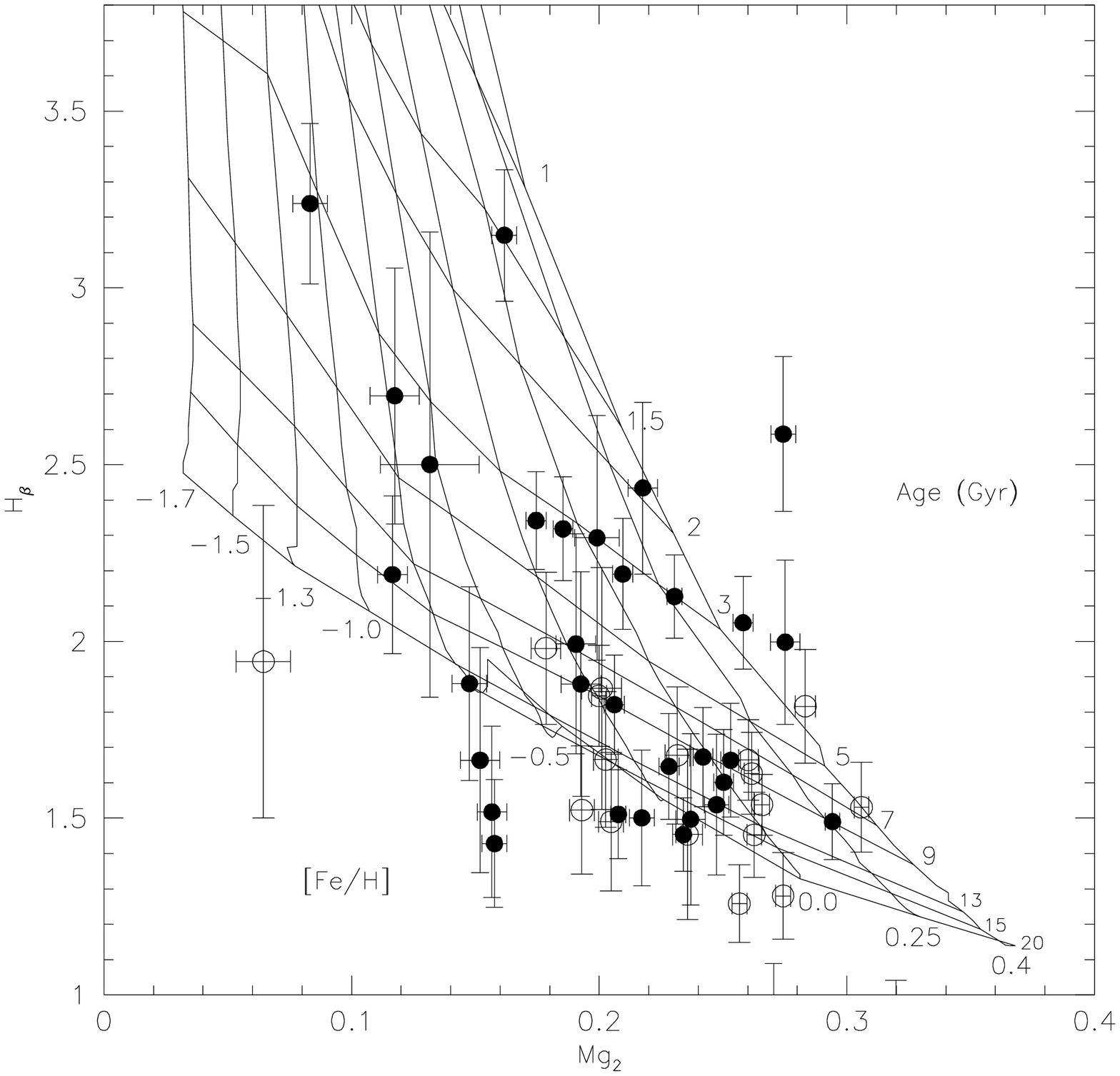}
\caption{$\rm H\beta$ versus $\rm Mg_2$ index strength for ellipticals
(empty circles) and S0 galaxies (filled circles).
Overplotted are models from Paper III (see text). Random errors
in the index measurements -- taking into account the variance in each pixel 
and the statistical propagation of errors (see Paper III) -- 
are shown. \label{fig1}}
\end{figure}

\clearpage 

\begin{figure}
\plotone{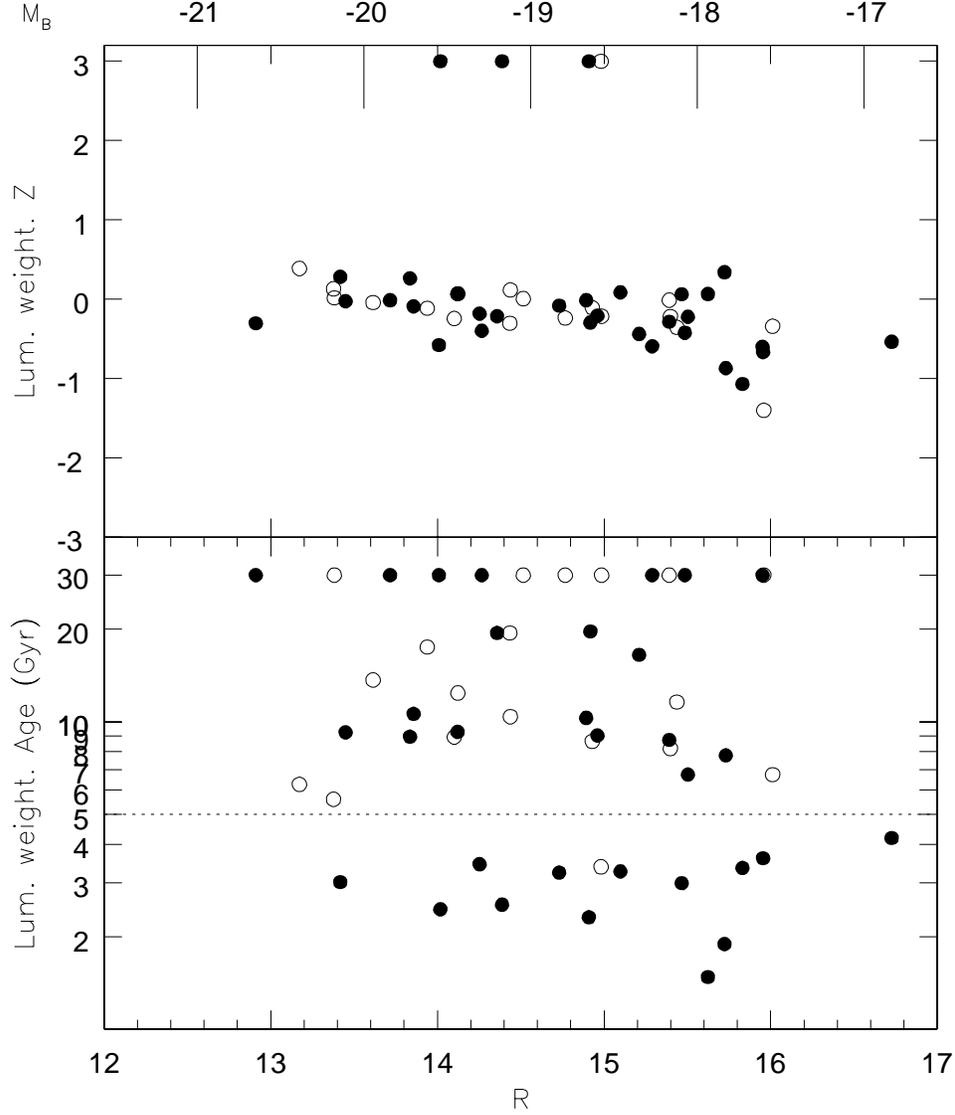}
\caption{Luminosity weighted metallicities (top) and ages (bottom)
of ellipticals (empty circles) and S0 galaxies (filled circles) as a function
of the R-band magnitude. The absolute B magnitudes shown on top of the plot
have been found for B-R=1.6.
The observed R band magnitudes are aperture magnitudes over a radius
3 times the Kron radius and are taken from \citet[Paper I]{kom01}.
Datapoints falling outside (on the right-hand side)
of the model grid in Fig.~1 have been arbitrarily
assigned a Z=3 in this plot. Similarly, an age = 30 Gyr was 
recorded when the datapoints lied below the model grid (see 
Paper III for a discussion of the mismatch between 
the observations and the model grid).
The metallicity-luminosity relation of the top panel for all
galaxies in this sample (Es+S0s) is $Z=4.555-0.318 \times R$, $\sigma=0.238$
(excluding the points lying at Z=3).
\label{fig2}}
\end{figure}

\clearpage 

\begin{figure}
\plotone{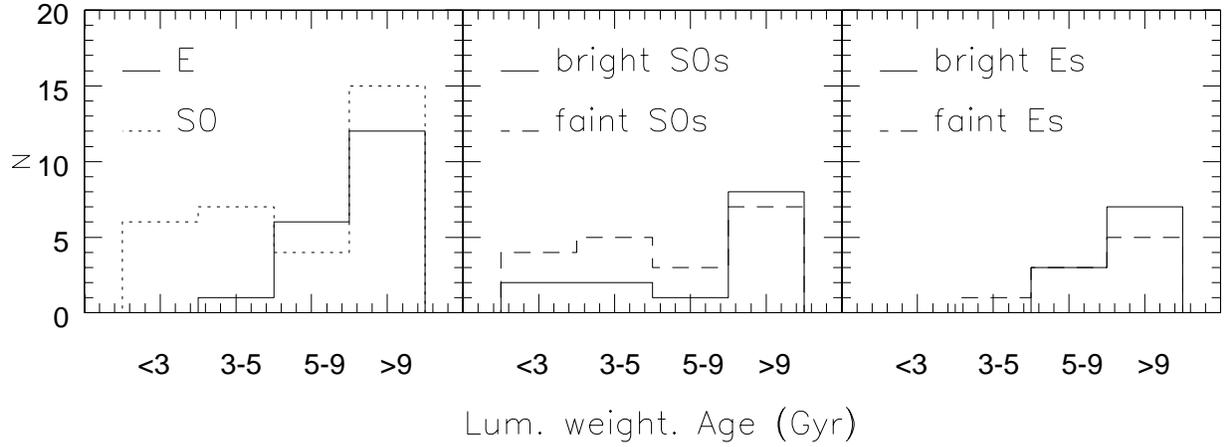}
\caption{Luminosity-weighted age distributions of ellipticals
and S0 galaxies (left panel). In the middle and right panels
the brightest ($R<14.5$) and faintest ($R>14.5$) subsets are plotted
separately.\label{fig3}}
\end{figure}

\begin{figure}
\plotone{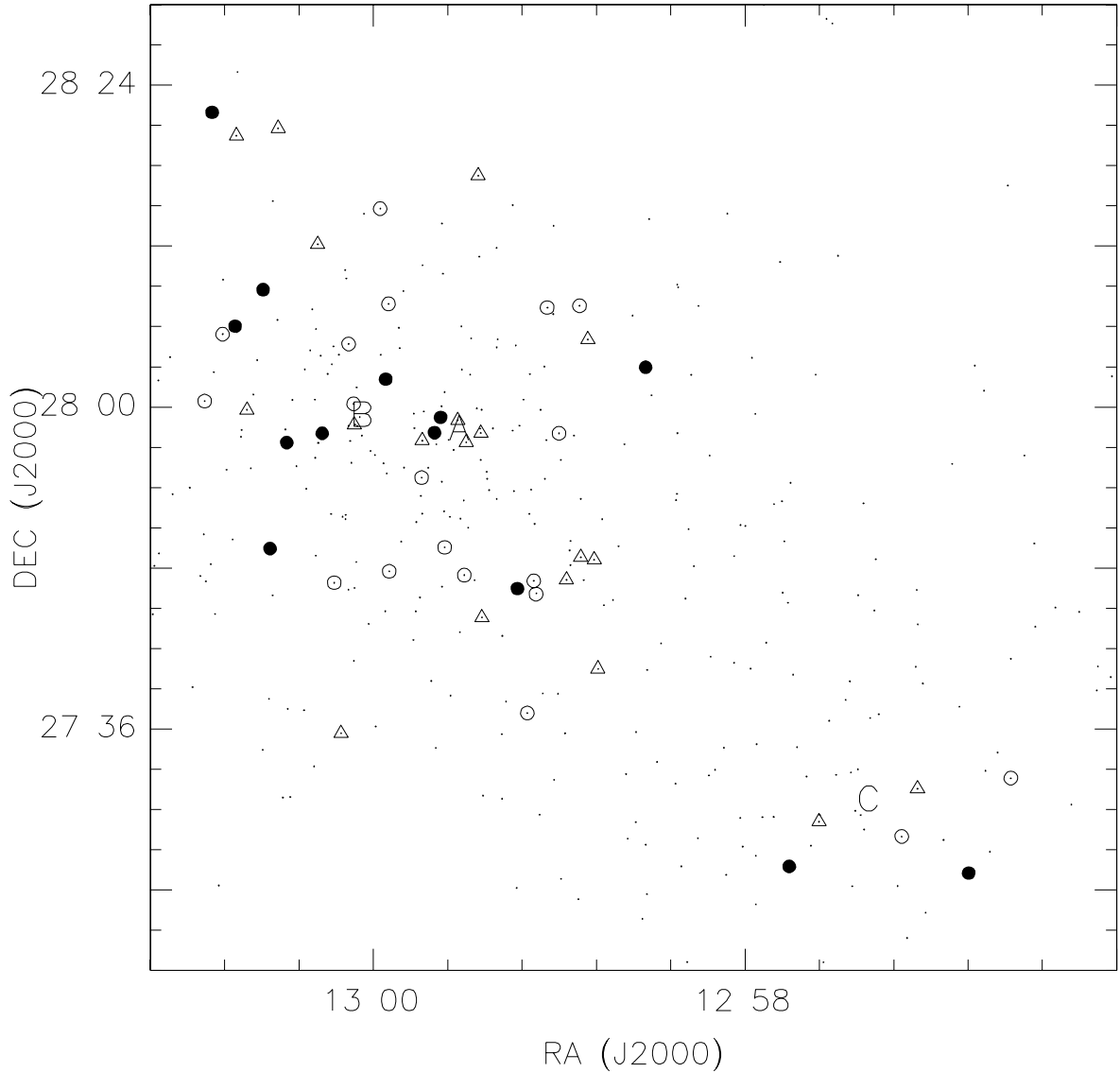}
\caption{Positions in the sky of the early-type galaxies.
Here we only show the area where there are galaxies in common between
our and Dressler's sample. 
S0 galaxies with luminosity-weighted ages $< 5$ Gyr are plotted
as filled circles,
while older S0s are represented by empty triangles. 
Empty circles are ellipticals.
Small dots represent galaxies known to be members of the cluster
from \citet{col96}. 
The positions of the two central
dominant galaxies (NGC4874 and NGC4889) are indicated
by the symbols ``A'' and ``B'', respectively. The cD galaxy in the
South-West group (NGC4839) is identified with the symbol ``C''.\label{fig4}}
\end{figure}

\clearpage

\end{document}